\documentclass{svjour3}
\smartqed  
\usepackage{graphicx}

\begin{document}

\title{Weakly bound states of polar molecules in
bilayers\footnote{Presented at the 21st European Conference on Few-Body
Problems in Physics, Salamanca, Spain, 30 August - 3 September 2010}
}


\author{D.V.~Fedorov \and J.R.~Armstrong \and N.T.~Zinner \and
	A.S.~Jensen}


\institute{
D.V.~Fedorov \at Aarhus University, 8000 Aarhus C, Denmark
\and J.R.~Armstrong \at Aarhus University, 8000 Aarhus C, Denmark
\and N.T.~Zinner \at Aarhus University, 8000 Aarhus C, Denmark
\and A.S.~Jensen \at Aarhus University, 8000 Aarhus C, Denmark
}

\date{Received: date / Accepted: date}

\maketitle

\begin{abstract}
We investigate a system of two polarized molecules in a layered
trap. The molecules reside in adjacent layers and interact purely via the
dipole-dipole interaction.  We determine the properties of the ground state
of the system as a function of the dipole moment and polarization angle.
A bound state is always present in the system and in the weak binding
limit the bound state extends to a very large distance and shows universal
behavior.
	\keywords{dipole-dipole interaction  \and 2D geometry \and
weakly bound states \and universality}
	\end{abstract}

\section{Introduction} \label{intro}
Cold gases of polar molecules represent an interesting opportunity
to study the interplay of long- and short-range interactions in the
controllable environment of the magneto-optical trap~\cite{micheli}.
One of the possible configurations is an optical lattice
with a multi-layered stack of planar traps.

We consider a system of two polar molecules in adjacent layers.  The
unique property of the system is that the interaction between molecules
is pure dipole-dipole. This interaction has the integrated strength
equal to zero -- the attractive and repulsive parts of the potential are
of equal integrated strength.  In 2D-geometry such interaction always
supports a bound state~\cite{simon}. We investigate the energy and
root-mean-square radius of the ground state as a function of the dipole
moment and polarization angle. In particular we show that in the
weak binding limit the state shows universal behavior.

\section{The Schr\"{o}dinger equation for two dipoles in bilayer geometry}
\label{sec:1}
The system of two polar molecules in a bilayer geometry is described by
the Schr\"{o}dinger equation,
	\begin{equation} \label{eq-se}
\left[-\frac{\hbar^2}{2\mu}\left( \frac{\partial^2}{\partial
x^2}+ \frac{\partial^2}{\partial y^2} \right) +
W(x,y)\right]\psi(x,y)=E\psi(x,y) \;,
	\end{equation}
where $\mu$ is the reduced mass of the two molecules, $x$ and $y$ are the
relative coordinates in the layer plane, and $W$ is the dipole-dipole
potential,
	\begin{equation} \label{eq-w}
W(x,y)=D_2 \frac{x^2+y^2+b^2-3(x\cos\theta +
b\sin\theta)^2}{(x^2+y^2+b^2)^{5/2}} \;,
	\end{equation}
where $b$ is the distance between the layers, $\theta$ is the polarization
angle measured from the layer, and $D_2$ is the product of the dipole
moments of the molecules taken with the positive or negative sign if
the dipoles are correspondingly parallel or anti-parallel.

\begin{figure}
\centering
\input{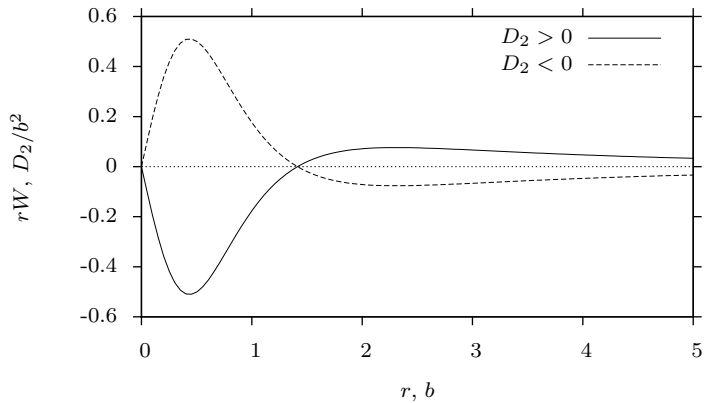}
\caption{The dipole-dipole potential~(\ref{eq-w})
in the rotationally symmetric case, $\theta=\pi/2$, as a function of
$r=\sqrt{x^2+y^2}$. The potential is multiplied by $r$ for better
visibility.}\label{fig-vr}
\end{figure}

The dipole-dipole potential has the integrated strength identically equal
to zero -- it has attractive and repulsive parts with equal integrated
strength, see figure~\ref{fig-vr}.  Such a potential is expected to always
have a bound state in a two-dimensional geometry~\cite{simon}.

\section{Ground state energy} \label{sec:gs}

	\begin{figure}[t]
\centering
  \input{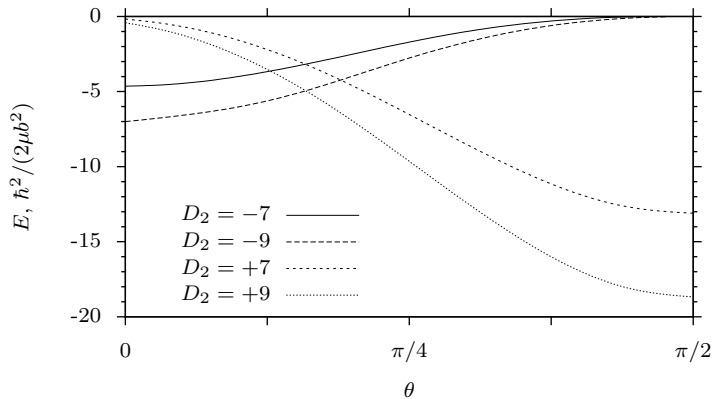}
\caption{Ground state energy of the system of two polar molecules
described by the Schr\"{o}dinger equation~(\ref{eq-se}) as a function
of the polarization angle $\theta$ for parallel, $D_2>0$, and
anti-parallel, $D_2<0$, orientation of the dipoles. $D_2$ is in units
of $\hbar^2b/(2\mu)$.}\label{fig-e} \end{figure}

We have calculated numerically the binding energy of the ground
state of the system of two polarized molecules for different dipole
moments as a function of the polarization angle for parallel
and anti-parallel orientation of the dipoles.  The results are shown
in figure~\ref{fig-e}.

For parallel dipoles the strongest binding is achieved when the
dipoles are perpendicular to the layer, $\theta=\pi/2$. The binding
decreases monotonically with increasing polarization angle. Approaching
$\theta=0$, the binding decreases significantly but still remains finite.
For anti-parallel dipoles the situation is opposite -- the binding is
strongest for $\theta=0$ and decreases towards $\theta=\pi/2$.  In both
cases a stronger dipole moment leads to stronger binding.

\section{Root-mean-square radius and universality in the weak binding limit}
\label{sec:rms}

	\begin{figure}[b]
	\centering
	\input{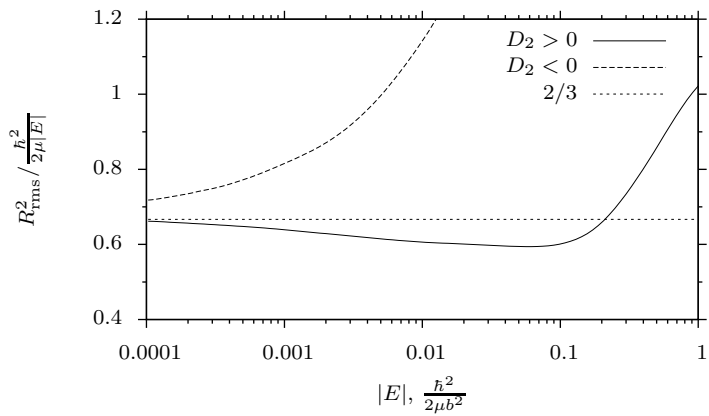}
	\caption{Root-mean-square radius $R_\mathrm{rms}$ of the system of two
polar molecules described by the Schr\"{o}dinger equation~(\ref{eq-se})
as a function of the binding energy $|E|$ for parallel, $D_2>0$, and
anti-parallel, $D_2<0$, orientation of the dipoles.  The binding energy
was varied by changing the dipole-dipole strength $D_2$ while the
polarization angle was fixed at $\theta=\pi/2$. The model independent
(universal) asymptotic value for vanishing binding is $2/3$.}
	\label{fig-rms}
	\end{figure}

For short-range potentials in two dimensions the relation between the
root-mean-square radius, $R_\mathrm{rms}$, and the binding energy, $|E|$,
is universal in the limit of weak binding~\cite{nielsen},
	\begin{equation}
R_\mathrm{rms}^2 \stackrel{E\to 0}{\longrightarrow} {2\over
3}{\hbar^2\over 2\mu|E|}.
	\end{equation}

The dipole-dipole potential asymptotically decreases as $r^{-3}$, that
is faster than the centrifugal barrier, $r^{-2}$. It is thus of
a short-range type and must obey the universal relation.  Indeed the
numerical calculations show that the universal regime is approached at
$|E|\sim 0.001 \frac{\hbar^2}{2\mu b^2}$, see figure~\ref{fig-rms}.
For the parallel (anti-parallel) orientation of the dipoles the universal
limit is reached from below (above) reflecting the geometry of the
dipole-dipole potential for the two configurations -- either
a barrier or a core, as illustrated in figure~\ref{fig-vr}.

\section{Conclusion}
\label{sec:con}
We have investigated the ground state of a system of two polar molecules
in a bilayer geometry. The ground state is always bound with the
binding energy strongly dependent on the polarization angle. In the
weak binding limit the system exhibits the universal relation between
the root-mean-square radius and the binding energy.

\end{document}